\documentstyle[amsfonts,aps,epsf]{revtex}
\newcommand{\beq}{\begin{equation}}
\newcommand{\eeq}{\end{equation}}
\newcommand{\beqa}{\begin{eqnarray}}
\newcommand{\eeqa}{\end{eqnarray}}
\newcommand{\non}{\nonumber}
\begin{document}

\title{Role of saddles in mean-field dynamics above the glass transition}

\author{Andrea Cavagna$^\star$, Irene Giardina$^\dagger$, 
and Giorgio Parisi$^\ddagger$}

\address{$^\star$ 
Department of Physics and Astronomy, The University, Manchester, 
M13 9PL, United Kingdom}
\address{$^\dagger$ 
Service de Physique Th\`eorique, CEA Saclay, 
91191 Gif-sur-Yvette, France}
\address{$^\ddagger$
Dipartimento di Fisica, Unit\`a INFM and Sezione INFN
Universit\`a di Roma ``La Sapienza'', 00185 Roma, Italy}

\date{April 27, 2001}

\maketitle

\begin{abstract}
Recent numerical developments in the study of glassy systems have shown
that it is possible to give a purely geometric interpretation of the dynamic
glass transition by considering the properties of unstable saddle points of the 
energy. 
Here we further develop this program in the context of a mean-field 
model, by analytically 
studying the properties of the closest saddle point to 
an equilibrium configuration of the system. 
We prove that when the glass transition is approached the energy 
of the closest saddle goes to the threshold energy, defined as the energy 
level below which the degree of instability of the typical stationary points 
vanishes.
Moreover, we show that the distance between a typical  
equilibrium configuration and the closest saddle is always 
very small and that, surprisingly, it is almost independent of 
the temperature.
\end{abstract}

\vskip 0.5 truecm


The glass transition occurs when the relaxation time of a substance increases 
upon cooling of many orders of magnitude in a very narrow interval of temperature, 
without the onset of any crystalline order.
Even though dramatic changes in the mechanical properties of the sample 
occur, it is impossible to define a strict transition temperature, 
because the dynamic process leading to the glassy phase is continuous, 
albeit very sudden.
This is indeed one of the most tricky points in the study of glassy systems: 
the so-called glass transition cannot actually  be associated to the genuine 
divergence of any dynamic or thermodynamic quantity. 
In fact, its very definition as a reference temperature is based upon convention: 
in supercooled liquids, it has been  agreed to fix the glass transition temperature 
$T_g$ at the point where the viscosity of the sample (that is one of the macroscopic 
manifestation of the relaxation time) is of order $10^{13}P$.

On the other hand, in the case of fragile super-cooled liquids \cite{angell}
there is at least another important value of the temperature which 
is  useful  to describe and interpret experimental data, that is
the temperature $T_{MCT}$ where Mode Coupling Theory (MCT) 
locates a purely dynamic transition \cite{mct}.
Such a transition is spurious, since what is observed 
in real experiments and simulations is just a dynamical crossover  
from a diffusive regime to an Arrhenius (or super Arrhenius) one.
However, MCT describes well the dynamics of fragile liquids for $T>T_{MCT}$
and, even though the MCT transition is smeared 
out in reality, still the MCT temperature remains a 
meaningful reference value marking the border between 
purely diffusive and activated slow dynamics \cite{angell2}.

The lack of a strict dynamic transition is not common to
all glassy systems. It has been discovered in the past \cite{tirumma}
that some {\it mean-field} models for 
spin-glasses display a phenomenology quite similar to the one of real 
structural glasses and supercooled liquids, but for a notable difference: 
in these mean-field systems there is a true divergence of the relaxation
time, with no associated thermodynamic anomaly. This fact
makes the definition of a dynamic critical temperature $T_d$ completely unambiguous 
for these models. Besides, the dynamical equations which describe 
the behaviour of  these systems above $T_d$ coincide with those obtained 
by MCT \cite{revmct}.  Thus,  for these models 
MCT is exact, and  $T_d$ therefore coincides with $T_{MCT}$. 
The most deeply studied among these systems is the $p$-spin spherical 
model \cite{crisa1,crisa2,ck,kpv,crisa3,noi-jpa1,noi-prb1}, 
henceforth indicated as $p$SM. 

A key feature of the $p$SM is the possibility to explain the dynamic
glass transition at $T_d$ as the result of a purely geometric transition
taking place in the energy landscape of the system \cite{ck,kpv,crisa3}. 
At a given energy density, called {\it threshold energy} $E_{th}$, there is a 
qualitative
change in the stability properties of the landscape: below $E_{th}$
and down to the ground state energy $E_0$
minima dominate, whereas above $E_{th}$ unstable saddles are the 
most numerous stationary points of the Hamiltonian. It can be proved that
in such a system the dynamic glass transition occurs when 
the equilibrium energy  density becomes equal to the energy density of 
the threshold states at that temperature \cite{ck,kpv}. In other 
words, in the $p$SM the dynamic glass transition at $T_d$ and the geometric 
transition at $E_{th}$ are essentially two faces of the same phenomenon.
Due to this fact, the structure and properties of unstable stationary
points in the $p$SM have been the object of a number of investigations 
in recent years \cite{laloux,noi-prb1,silvio}.

Infinite lifetime metastable states cannot exist in finite dimensional
systems, and therefore we cannot expect to find a divergence of the 
relaxation time with no associated thermodynamic transition in 
non-mean-field models.
However, a strict geometric transition at a threshold energy 
may very well occur also in   more realistic systems, such as
supercooled liquids, even if its 
dynamical counterpart is smeared out by the finite dimensional nature of the 
system (i.e. by the finite lifetime of the threshold minima).
If this were true, the dynamic crossover at $T_{MCT}$, which in 
supercooled liquids marks the onset of activated glassy dynamics, 
would actually 
be the manifestation of a more fundamental and sharply defined 
geometric transition occurring at a certain critical threshold energy. 

This scenario has been numerically investigated very recently in 
\cite{sad,ruocco} for various Lennard-Jones (LJ) systems, 
and from a more speculative point of view in \cite{bradipo}. 
In particular, in \cite{sad} a well defined threshold potential 
energy has been located and associated to the onset of glassy 
dynamics in the system. 
Moreover, it has been directly shown in \cite{kcl} that the energy landscape of 
LJ models and that of the $p$SM are indeed very similar. 
These studies seem therefore to confirm the idea that, even in 
realistic systems, the dynamic crossover observed at $T_{MCT}$
is the consequence of the sharp change in the topology 
of the energy landscape at the threshold energy.

The approaches of \cite{sad,ruocco,bradipo} all have as a vital starting 
point the assumption that the time evolution of the system in the 
phase space is in some way influenced by the nearby saddle points 
of the potential energy. 
However, this fact has not been directly proved, and the circumstantial 
evidences are mainly of a numerical nature. 
In particular, it may be objected that the trajectory of the 
system at equilibrium is never even {\it close} to saddles, especially above 
the glass transition, where it may be argued that free diffusion in the phase
space implies that the energy landscape and its stationary points are completely 
irrelevant (for the relevance of saddles in zero temperature dynamics see 
\cite{laloux}). 
Even in the $p$SM there was up to now little evidence of any {\it direct} 
connection between dynamics of the system equilibrated above $T_d$ and saddle points 
of the energy above $E_{th}$ (see, however, the approach of \cite{silvio}).
Furthermore, even assuming that the dynamic trajectory stays somewhat close 
to saddles, it remains to be directly demonstrated  
that such objects do play a role in the transition. 
More precisely, one should prove  that the properties 
of these supposedly close saddles indeed display some anomaly 
at the dynamic transition.

The aim of the present work is therefore to analytically investigate
what is the role of saddles in the equilibrium dynamics of the $p$SM above $T_d$. 
In order to do this we will introduce a tool,
which allows for the exact location of the closest saddle points to an 
equilibrium configuration at temperature $T$. 
In this way we will be able to study how the properties of these closest 
saddles vary with the temperature when the dynamic transition is approached, 
thus answering some of the questions raised above.

The Hamiltonian of the $p$SM is given by,
\beq
H= \sum_{i_1<\cdots<i_p}^N J_{i_1\cdots i_p} \tau_{i_1} \cdots \tau_{i_p} =
\frac{1}{p!}\sum_{i_1<\cdots<i_p}^N J_{i_1\cdots i_p}  \tau_{i_1}
\cdots \tau_{i_p} + O(1/N) \ ,
\eeq
where the spins satisfy the spherical constraint $\sum_i\tau_i^2=N$. 
The quenched couplings
$J_{i_1\cdots i_p}$ are Gaussian distributed random variables with variance
${J^2}=p!/2N^{p-1}$. 
By means of the Lagrange method, we can find the stationary points of 
the Hamiltonian on the sphere and therefore 
write the equations satisfied by the saddle points 
of $H$ with energy density $E$,
\beq
\frac{1}{p!}\sum_{i_2\cdots i_p}^N J_{k,i_2 \cdots i_p} \tau_{i_2}
\cdots  \tau_{i_p} - E\tau_k=0 \quad \quad , \quad \quad k=1,\dots,N \ .
\label{staz}
\eeq
In general, the number ${\cal N}(E)$ of solutions of equations 
(\ref{staz}) with energy density $E$ 
is exponentially large in the size of the system $N$. 
Thus, the quantity which is normally computed is the {\it complexity}
(or configurational entropy), defined as the logarithmic density of this
number, $\Sigma(E)=\frac{1}{N}\log {\cal N}(E)$. 
The  nature of the saddle points of $H$
is in principle  not only specified by their energy density $E$, but also 
by their instability index $K$, that is the number of negative 
eigenvalues of the Hessian matrix. 
However, previous studies of the $p$SM have shown that there is a well 
defined relation between energy density and index $K(E)$, 
and that {\it at any fixed energy level $E$} 
only stationary points with index $K(E)$ dominates the 
energy landscape in the thermodynamic limit \cite{kpv,noi-prb1,juanpe}.
Therefore, by fixing the energy density of a saddle point to $E$, we are 
automatically fixing its index to $K(E)$. 
A crucial feature of the $p$SM is that the typical saddles index is 
extensive, $K=O(N)$, as long
as the energy density is above a value called {\it threshold}, $E_{th}$, while
$K=0$ for $E\leq E_{th}$. This means that minima dominate over saddles
below the threshold, while saddles of index $K(E)>0$ are the most numerous 
stationary points for $E>E_{th}$. In this sense, we can say that at $E_{th}$
a geometric transition takes place.
More precisely, if we introduce the index density $k=K/N$, we have,
\beqa
k(E)&=&\frac{p}{\pi(p-1)} \left[
\arctan \left(-\frac{\sqrt{E_{th}^2-E^2}}{E}\right)
+\frac{E}{4} \sqrt{E_{th}^2-E^2}
\right] 
\quad , \quad E \geq E_{th} \non \\
k(E)&=& 0 \phantom{
\arctan \left(-\frac{\sqrt{E_{th}^2-E^2}}{E}\right)
+\frac{1}{4\pi}E \sqrt{E_{th}^2-E^2}pppppppp} 
\quad , \quad E \leq E_{th} \ .
\eeqa
Note that $k(E)$ is a monotonically increasing function of the energy $E$.
Remarkably, when  
the equilibrium energy density of the system becomes equal to the internal 
energy density of the threshold minima  the system undergoes a dynamic glass
transition, which we will indicate with $T_d$.
To better specify this statement, we have to distinguish between the 
{\it bare} energy $E$ of a minimum, and its internal energy $U(T)$,
that is the energy of a system equilibrated in that minimum at 
temperature $T$ 
\footnote{ In other terms, if $m_i$ is the local magnetization of the
system equilibrated in the minimum, and $q=1/N \sum_i m_i^2$ is the
{\it self-overlap} (i.e. the magnetization norm, 
which is related to thermal fluctuations), 
we can define a bare magnetization as ${\hat
m}_i=\frac{m_i}{\sqrt{q}}$. The bare energy density is then given by
$E=\frac{1}{N} H({\hat m}_i)$.}. 
Of course, the quantity $U(T)$ is equal to the bare energy $E$ plus a 
vibrational contribution due to thermal fluctuations. 
At the dynamic glass transition
$T_d$ we have that $U_{eq}(T_d)=U_{th}(T_d)$, where $U_{eq}(T)$ is the global 
equilibrium energy density of the system, and $U_{th}(T_d)=E_{th}+vibrations$.

Our aim in this paper is to analyze the structure of the 
saddle points around an equilibrium 
configuration thermalized at temperature $T>T_d$. We therefore need
a notion of distance to give a meaning to this statement. 
Given two configurations $\sigma$ and $\tau$  we define a 
co-distance, or {\it overlap}, $q_{\sigma\tau}$, as, 
\[
q_{\sigma\tau}=\frac{1}{N}\sum_i^N \sigma_i \tau_i \ .
\] 
Similar configurations have $q\sim 1$, while different ones have $q\sim 0$.
Our strategy will be to fix a reference equilibrium configuration
$\sigma$ and compute the complexity of the saddles points $\tau$ close to it
as a function of their overlap $q_{\sigma\tau}$. The value of the 
overlap where  this quantity goes to zero will give the distance
of the closest stationary points to $\sigma$. Indeed, for larger overlaps, i.e. 
smaller distances, a negative complexity indicates a vanishing probability
of finding a stationary point.

In order to do this we have to calculate how many saddles
$\tau$, with a given energy $E$, happen to have an overlap $q$ with a
reference equilibrium configuration $\sigma$.
Clearly, this number formally depends on $\sigma$ itself and on the 
disorder $J$. However, as always done in similar calculations
\cite{v2,v3,noi-jpa1}, we can assume that in the thermodynamic
limit $N\to\infty$ this quantity is self-averaging with respect 
to the distribution of $\sigma$ and $J$, and therefore we can 
average it over the Gaussian 
distribution of the disorder (indicated with a bar) 
and over the equilibrium distribution of $\sigma$ at temperature
$T$. In this way we can define the saddles complexity as,
\[
\Sigma_s(q,E,\beta) \equiv \frac{1}{N}\overline{
\int \frac{D\sigma}{Z(\beta)}\ e^{-\beta H(\sigma)}} \ \times
\]
\beq
\overline{\log \int D\tau \ \prod_k
\delta\left(\frac{1}{p!}J_{k, i_2\cdots i_p}\tau_{i_2}\cdots \tau_{i_p}
-E\tau_i\right)
\ \left|\det\left(\frac{1}{p!}J_{k,l,i_3 \cdots i_p}\tau_{i_3}\cdots \tau_{i_p}
-E\delta_{kl}\right)\right| \
\delta(q-q_{\sigma\tau}) }\ ,
\label{lei}
\eeq
with,
\[
Z(\beta)=\int d\sigma \, e^{-\beta H(\sigma)} \ ,
\]
and where integration is carried out over spherical configurations only.
To understand equation (\ref{lei}) it is convenient to read it 
right-to-left: first, under the $\tau$ integral, we calculate
using the standard method of \cite{braymoore}
the number of solutions of equations (\ref{staz}), putting an 
extra constraint on the overlap they must have with $\sigma$.
Second, we take the logarithm of this quantity, and we average it 
over the equilibrium distribution of $\sigma$. Finally, we average
everything over the disorder $J$.
As we can see, $\Sigma_s$ depends on the temperature
$T=1/\beta$ at which the reference configuration $\sigma$ is equilibrated, 
on the energy $E$ of the saddles $\tau$ we are counting, 
and finally on the overlap $q$ between $\sigma$ and $\tau$. 
We stress that, by construction, $\sigma$ is an 
independent equilibrium configuration, irrespective of the energy and
the distance of the saddle $\tau$. 

In order to perform the averages
in (\ref{lei}) it is convenient to use the replica method, writing
\beqa
Z^{-1} &=& \lim_{n\to 0}\ Z^{n-1} \non \\
\langle \log (\cdot)\rangle &=& \lim_{m\to 0}\ 
\frac{1}{m} \log\langle (\cdot)^m\rangle \ .
\eeqa 
In this way we have,
\[
\Sigma_s(q,E,\beta) = \lim_{n,m\to 0} \frac{1}{Nm}\log  \overline{
\int  D\sigma_a\ D\tau_\alpha \ e^{-\beta \sum_a H(\sigma_a)} } \ \times 
\]
\beq
\overline{
\prod_{k\alpha}
\delta\left(\frac{1}{p!}J_{k,i_2\cdots i_p}\tau^\alpha_{i_2}\cdots 
\tau^\alpha_{i_p}-E\tau^\alpha_i\right)
\ \left|\det\left(\frac{1}{p!}J_{k,l,i_3\cdots i_p}\tau^\alpha_{i_3}\cdots
\tau^{\alpha}_{i_p}-E\delta_{kl}\right)\right| \
\delta(q-q_{\sigma_1 \tau_\alpha}) }  \ ,
\label{zuppa}
\eeq
\noindent
with $a=1,\dots,n$ and $\alpha=1,\dots,m$. 
The explicit calculation of 
$\Sigma_s$ from equation (\ref{zuppa}) can be performed 
by using the standard tools of the replica method: variational
parameters are introduced and the integrals are evaluated exactly
in the limit $N\to\infty$ by means of 
the steepest descent method. Of course, 
this is possible thanks to the mean-field nature of the model.
Here, we will skip most of the details and just 
state the final result. The interested reader may refer to 
\cite{noi-jpa1}, where a technically similar calculation is performed.
The full expression for the saddle complexity is: 
\beqa
\Sigma_s(q,E,\beta)= &&
\frac{1}{2} + 
\frac{p E^2}{2(p-1)} +
\frac{1}{2}\log\left(\frac{p-1}{2p}\right) +
\frac{1}{4p}(x_1-x_0 r^{p-1}) + 
\frac{1}{2} \beta q^{p-1} w + 
Ey_1 + 
\non  \\
&&
\frac{p-1}{4p}\left(y_1^2-y_0^2r^{p-2}\right) +
\frac{1}{2} \log \Omega_1 + 
\frac{\Omega_1}{2\Omega_2} 
+\frac{1}{2} \left(\frac{r-q^2}{1-r}\right) \ ,
\eeqa
with, 
\beqa
\Omega_1 &=& (x_1-x_0)(1-r) + (y_1-y_2)^2 \non \\
\Omega_2 &=& (x_0+w^2)(1-r) + 
(y_1-y_0)[2(y_0-qw) - (r-q^2)(y_1-y_0)/(1-r)] . 
\eeqa
As customary in the context of the replica method, the set of 
variational parameters ${\bf x}=(x_0, x_1, y_0, y_1, w, r)$ is 
fixed be means of the steepest descent equations 
$\partial \Sigma_s/\partial{\bf x}=0$, which we have solved 
numerically. 
We remark that the expression above for $\Sigma_s$ 
is only valid in the regime $E \geq E_{th}$. In showing the
results we will assume $p=3$.

First of all, we 
are interested in studying the behaviour of $\Sigma_s(q,E,T)$ as a function 
of $q$, at fixed $E$ and $T$.
In this way we can define an overlap 
$q_0(E,T)$ where $\Sigma_s$ goes to zero: this overlap
gives the distance of the closest saddle with energy $E$ to
an equilibrium configuration at temperature $T$. 
In Figure 1  we plot $\Sigma_s$ as a function of the overlap $q$,
for $T=T_d$ and $E=E_{th}$. At this temperature
many properties of equilibrium landscape are known and an interpretation of
the results is therefore much simpler.   
At $T_d$  the system equilibrates inside a threshold state   
with bare energy density $E_{th}$ and self-overlap (i.e. largeness) $q_{th}$ 
\cite{kpv}. In naive terms we can then imagine that
our typical equilibrium configuration  $\sigma$
lies in a well whose largeness is given by $q_{th}$ and whose 
bottom is at energy density $E_{th}$. In this  
case it is evident that the closest stationary  point to $\sigma$ is precisely
the bottom of the well. 
The point where the complexity goes 
to zero  must therefore give the overlap between the center of the
threshold minimum $\tau$ and one of its typical equilibrium 
configurations $\sigma$. 
This overlap can be easily computed by noticing that 
$q_{\sigma\tau}=\frac{1}{N}\sum_i \langle \sigma_i \tau_i \rangle=
\frac{1}{N}\sum_i m_i \tau_i$, where the thermal average is
restricted to a threshold state, and $m_i$ indicates the local
magnetization of that state. In the $p$SM this local magnetization
can be expressed directly in terms of the  
well  minimum as $m_i=\sqrt{q_{th}}\tau_i$ \cite{kpv} (see also 
footnote 1) and we immediately get 
$q_{\sigma\tau}=\sqrt{q_{th}}$.
Consistently with this result we find,
\beq
q_0(E_{th},T_d)=\sqrt{q_{th}} \ .
\eeq
This result can be appreciated in Figure 1.
\begin{figure}
\begin{center}
\leavevmode
\epsfxsize=3in
\epsffile{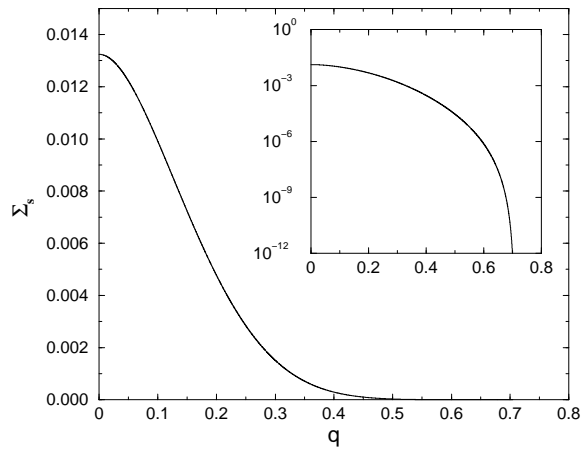}
\caption{Saddles complexity as a function of the overlap $q$, for $E=E_{th}$ and
$T=T_d$. Inset: same curve in linear-log scale. The complexity goes to zero at
$q_0=\sqrt{q_{th}}=0.71$, for $p=3$.}
\end{center}
\end{figure}
\noindent
Moreover, a careful analysis of $\Sigma_s$ for $q\sim q_0$ shows that,
\beq
\Sigma_s(q,E_{th},T_d) \sim (q-q_0)^5 
\quad ,  \quad q\sim q_0
\ .
\eeq
Note that the exponent is the same as found in \cite{noi-jpa1}
for the approach to zero of the constrained complexity
of threshold Thouless-Anderson-Palmer (TAP) solutions. This is a 
consistency check for the present calculation.

In the light of our original aim to find the closest saddles, it is 
interesting to plot the value of $q_0$ as a function 
of the energy density $E$ of the saddles we are counting. 
We expect this curve to have a maximum at a value $E_s(T)$
corresponding to the energy of the closest, and thus the most relevant, 
saddles. Accordingly, in Figure 2 we plot $q_0(E)$ as a function of $E$ 
for  different values of the temperature.
In particular, the full curve  
represents $q_0(E)$ for $T=T_d$: as we see, this is a steadily 
decreasing curve having its maximum at the threshold energy. 
This means that, as previously said,  the 
closest stationary point to an equilibrium configuration at the
glass transition is a threshold minimum, $E_s(T_d)=E_{th}$.
However, if we now increase the temperature $T$ of the equilibrium configuration,
we expect the energy of the closest saddle to increase as well,
together with its instability index. In other words, the higher the
temperature of $\sigma$, the higher will be the energy, and thus the degree
of instability, of the closest possible saddle (we remind that $K(E)$ is a 
monotonic increasing function of $E$, with $K(E_{th})=0$).
This hypothesis is confirmed in Figure 2: the maximum of these curves moves to the right as
the temperature is increased, disclosing a well defined relation $E_s(T)$, which
we will analyze carefully later. For the moment, let us note that a further 
consistency check of our calculation is that for $T=\infty$
we find $E_s=0$ (see Figure 2). At very high temperatures the equilibrium configuration 
$\sigma$ is just a random configuration of the system, therefore the closest 
stationary points to it will be the most numerous ones in absolute terms. In
the $p$SM it can be proved that the most numerous saddles have $E=0$ and 
$K=N/2$ \cite{kpv,juanpe}. 
\begin{figure}
\begin{center}
\leavevmode
\epsfxsize=3in
\epsffile{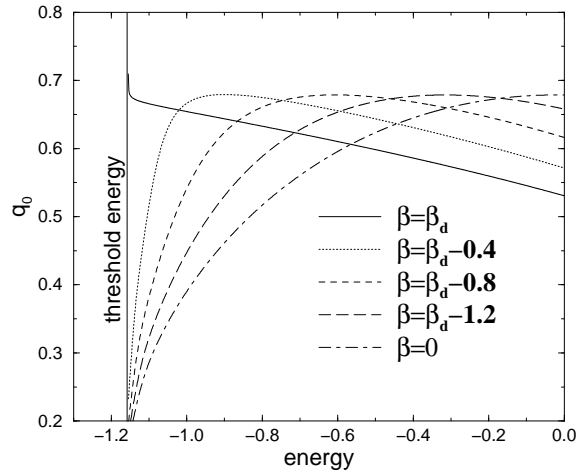}
\caption{The overlap $q_0$ where the saddles complexity goes to zero as a function 
of the energy $E$ of the saddles, at five different values of the temperature above $T_d$.
The full vertical line marks the value of the threshold energy $E_{th}$.}
\end{center}
\end{figure}
It is interesting to study the behaviour of the overlap $q_s$ where 
the curves $q_0(E)$ have their maximum. This overlap is a measure of
the closeness of an equilibrium configuration $\sigma$ to its nearest 
saddle point $\tau$.
What is surprising of Figure 2, is that by varying the temperature 
the value of $q_s$ is almost constant. 
To better investigate this point we plot in Figure 3
$q_s$ as a function of $\beta$. We can see that $q_s$ is practically always 
constant, but for $\beta\sim\beta_d$, where it sharply jumps to $\sqrt{q_{th}}$. 
This fact means that {\it the distance between an equilibrium configuration 
and its closest saddles is almost independent of the temperature}. This value
of the overlap is $q_s\sim 0.68$, which is indeed quite high, being comparable 
to the overlap between equilibrium configurations and bottom of the minima
below the glass transition (see, for example, Figure 1).
This result answers one of the main questions raised in the introduction: 
above the dynamic glass transition, 
the equilibrium trajectory indeed stays always very close to {\it unstable} 
stationary points of the Hamiltonian, exactly as below $T_d$ it stays close 
to {\it stable} minima. On the other hand, it is clear from 
Figure 2 that for $T>T_d$ the closest minima (i.e. the ones with $E=E_{th}$)  
are very far from the dynamic trajectory. 
In this sense, it is justified to 
say that the equilibrium dynamics of the system above the
glass transition may be described as an evolution among the neighborhoods 
of saddle points, rather than among basins of the minima
\cite{bradipo}.
\begin{figure}
\begin{center}
\leavevmode
\epsfxsize=3in
\epsffile{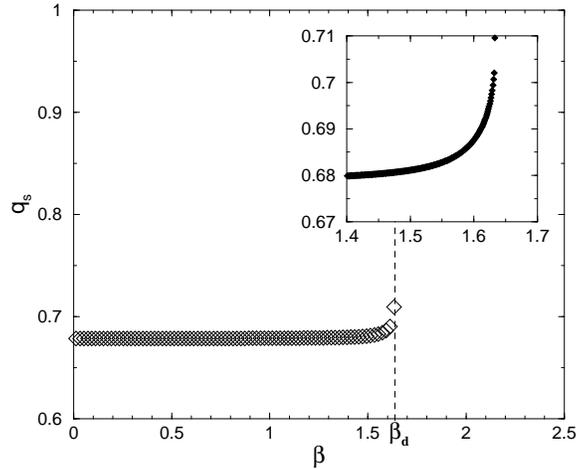}
\caption{The overlap $q_s$ of the closest saddles to an equilibrium configuration at temperature
$T$, as a function of $\beta=1/T$. Inset: enlargement of the same plot for $\beta\sim\beta_d$.}
\end{center}
\end{figure}
The distance of the closest saddle does not change with $T$, 
but as we have seen, the  energy density $E_s$ does. 
In Figure 4 we plot this energy  
as a function of the inverse temperature and compare it with the 
equilibrium energy density of the system above the glass transition, i.e. 
$U_{eq}(T)=-\beta/2$ \cite{crisa1}. 
First of all we note that the energy of the saddle
is always smaller than the energy of the equilibrium configuration, 
despite the two objects being so close in the phase space. 
This fact has been already noted in the context of a numerical study 
of a Lennard-Jones system in \cite{ruocco}. It is tempting to interpret
the difference $U_{eq}(T)-E_s(T)$ as a pseudo-vibrational 
contribution of saddles, due to the fact that, even though $K>0$, the 
largest part of the Hessian eigenvalues is positive, as long as 
$E_{th}<E<0$. 
Unstable saddles are not trapping object, of course,
but they may have a substantially long life-time provided that 
$K$ is small enough. This phenomenon is at the 
basis of the  pseudo-vibrational contribution of saddles.

This last hypothesis is supported by another interesting result 
we find, that is,
\beq
E_s(T) \to E_{th} \quad , \quad T \to T_d \ .
\eeq
Therefore, the energy density of the closest saddles to 
an equilibrium configuration goes to the
threshold energy density at the dynamic glass transition. Clearly, at $T_d$
the difference between $E_s(T_d)$ and $U_{eq}(T_d)$ is given by the 
vibrational contribution of thermal fluctuations inside
threshold minima, which is of order $k_BT$.
When $T>T_d$ we see that the two curves continuously approach one 
another, as the saddles instability index $K$ increases with the 
energy. These results seem thus to suggest that, even though the system
is {\it not} confined into any given saddle point, the disproportion
between trapping and un-trapping directions, that is the fact 
that $K < N/2$,  is sufficient to produce a vibrational contribution
that we may broadly interpret as thermal fluctuations around saddles 
point.
\begin{figure}
\begin{center}
\leavevmode
\epsfxsize=3in
\epsffile{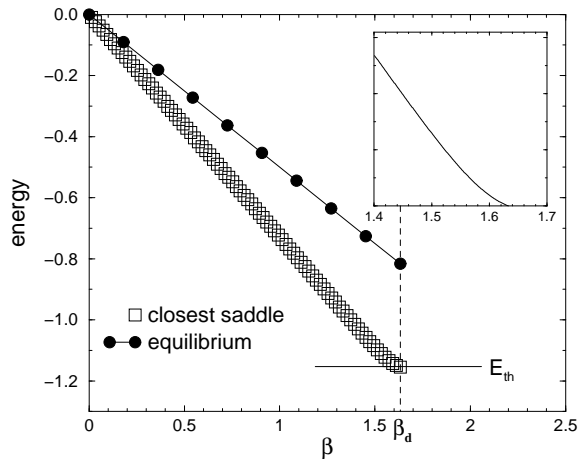}
\caption{The bare energy $E_s$ of the closest saddles compared to the equilibrium 
energy $U_{eq}$. Inset:  enlargement of the same plot for $\beta\sim\beta_d$. The slope of the 
curve changes in this regime.}
\end{center}
\end{figure}
A further support of this idea comes from the comparison of our results
with the approach developed in \cite{silvio}, where a dynamic description
of the $p$SM based on the concept of quasi-states is introduced. 
More specifically, for temperatures slightly above the dynamic transition 
$T_d$, quasi-states are related to critical points of the 
 TAP free energy (for a more detailed definition
see \cite{silvio}) and it can be shown that their global contribution
gives rise to the paramagnetic free energy of the system. 
Interestingly enough, we found that the bare energy density of these
quasi-states (defined as in footnote 1) is to a good degree of accuracy 
(within $1\%$) equal to the energy density $E_s$ of 
the closest saddles. This indicates 
that the quasi-states introduced in \cite{silvio} may be
interpreted as our closest saddles plus the pseudo-vibrational contribution
mentioned before. This interpretation  confirms the idea,
outlined in \cite{silvio} and made more explicit in the present paper,  that
above $T_d$ the paramagnetic state is made up of disjoint  quasi-states
around saddles. Equilibrium dynamics can thus be thought as evolution
from one quasi-state (i.e. one saddle and its own neighborhood) to
another one. 
A difference between the present approach and the one of \cite{silvio}
is that the pseudo-states of \cite{silvio} can be defined 
only very close to $T_d$, 
while, as we have seen, closest saddles exist at any temperature, although, of
course, we do not expect them to have any relevance for $T\gg T_d$.

Given the relation $k(E)$ between index and energy of the typical saddles,
we can introduce a temperature-dependent index by using the energy density
$E_s(T)$ of the closest saddles, namely $k(T)=k(E_s(T))$. Clearly, this
index vanishes at the dynamic transition temperature $T_d$, which is 
just another way of describing the geometric transition occurring at the
threshold energy $E_{th}$. Close to $T_d$ we find that $E_s$ is linear in 
$\beta$ and this, together with the analytic form of $k(E)$, implies that,
\beq
k(T) \propto (T-T_d)^{3/2} \quad \quad , \quad \quad T\sim T_d \ .
\eeq
Summarizing, in the context of the $p$SM 
we have calculated the complexity of the saddle points
at fixed overlap with a reference equilibrium configuration, above
the dynamic glass transition. In this way we were able to identify 
what are the energy and distance of the closest saddles at any given 
temperature. We found that the distance between equilibrium configuration
and closest saddle is almost independent of the temperature and is very
small. Moreover, the energy of the closest saddles intersects the 
threshold energy  at the dynamic glass transition. 
Finally, we interpreted the difference between 
equilibrium energy and energy of the closest saddle as a pseudo-vibrational
contribution due to the fact that a number of trapping directions larger
than the number of non-trapping ones may give rise to thermal fluctuations
around unstable saddle points. The present study supports the idea 
that dynamics in glassy systems for $T>T_d$ can
be described in terms of evolution in the phase space among 
the neighborhoods of unstable saddles and strengthens the hypothesis 
that the glass transition,
even in finite dimensional systems, is just the manifestation of the
topological transition between saddles and minima dominated regions 
of the phase space.

\vskip 0.5 truecm

It is a pleasure to thank Silvio Franz for an important discussion.
The work of AC was supported by EPSRC, under grant GR/L97698

\end{document}